**The Expresso Microarray Experiment Management System:
The Functional Genomics of Stress Responses in Loblolly Pine**

**Lenwood S. Heath[1], Naren Ramakrishnan[1], Ronald R. Sederoff[2], Ross W. Whetten[2], Boris I. Chevone[3], Craig A. Struble[1], Vincent Y. Jouenne[1], Dawei Chen[1], Leonel van Zyl[2], and Ruth G. Alscher[3]**

## Abstract:

Conception, design, and implementation of cDNA microarray experiments present a variety of bioinformatics challenges for biologists and computational scientists. The multiple stages of data acquisition and analysis have motivated the design of Expresso, a system for microarray experiment management. Salient aspects of Expresso include support for clone replication and randomized placement; automatic gridding, extraction of expression data from each spot, and quality monitoring; flexible methods of combining data from individual spots into information about clones and functional categories; and the use of inductive logic programming for higher-level data analysis and mining. The development of Expresso is occurring in parallel with several generations of microarray experiments aimed at elucidating genomic responses to drought stress in loblolly pine seedlings. The current experimental design incorporates 384 pine cDNAs replicated and randomly placed in two specific microarray layouts. We describe the design of Expresso as well as results of analysis with Expresso that suggest the importance of molecular chaperones and membrane transport proteins in mechanisms conferring successful adaptation to long-term drought stress.

# Introduction

Microarray technology makes possible the measurement of levels and patterns of gene expression important in growth, metabolism, development, behavior, and adaptation of living systems. A single microarray hybridization can provide information about the expression of hundreds or thousands of genes. The complexity of devising an appropriate microarray design, as well as the complexity of analyzing large amounts of resulting data, are well beyond human capabilities. The design of a microarray must address several interrelated issues: selection of cDNAs; replication of each cloned cDNA fragment; and placement of each replicate. A sophisticated computational approach to design is clearly needed. A typical hybridization yields two images with thousands of visible spots conveying information about relative levels of expression. However, the reliable capture of the expression information from images is an ill-specified problem that must be addressed with flexible techniques that adapt to the particular parameters of the experiment, as well as the presence of noise and artifacts in the images.

---

[1] Department of Computer Science, Virginia Tech, Blacksburg, VA 24061.
[2] Forest Biotechnology, College of Natural Resources, North Carolina State University, Raleigh, NC 27695.
[3] Department of Plant Pathology, Physiology, and Weed Science, Virginia Tech, Blacksburg, VA 24061.

## Choreography of Gene Expression Patterns

Microarrays provide a means of determining the transcript profiles of entire genomes under a given set of experimental conditions, where an entire genomic sequence is known. When sequencing is still in progress (e.g., maize, loblolly pine), there is the added potential for gene discovery. A cDNA microarray experiment increases our ability to find genes that are expressed in similar patterns over time or under particular conditions and to find patterns in expression data. Powerful tools for inferring function also come from the assignment of common responses to multiple environmental stresses. Much functional genomic information remains buried in the data already accumulated [Bard, 1999]. New computational tools for mining this hidden information are needed, as are tools for analyzing gene expression patterns to reveal unknown cellular regulatory networks and signal transduction pathways. Accomplishments brought about through the use of microarrays are already considerable [Epstein and Butow, 2000, Somerville and Somerville, 1999, Cushman and Bohnert, 2000]. Microarrays have been used to identify transcript profiles characteristic of particular human tumor types [Khan et al., 1999,Yang et al., 1999, Perou et al., 1999, Golub et al., 1999, Monni et al., 2001, Kannan et al., 2001]; cell cycle regulatory mechanisms [Cho et al., 1998, Chu et al., 1998]; salt stress [Kawasaki et al., 2001]; hypoxia [Gracey et al., 2001]; apoptosis [Brachat et al., 2000]; and oxidative stress responses [Jelinsky et al., 2000, Alexandre et al., 2001].

In plants, microarray experiments with Arabidopsis cDNAs [Schaffer et al., 2000] have revealed differential effects of wounding compared to insect feeding [Reymond et al., 2000]; induction of novel potential regulatory genes in response to nitrate [Wang et al., 2000]; differential responses to cold and drought stress [Seki et al., 2001]; potential regulatory sequences in developing seeds [White et al., 2000]; and differential expression among leaves, flowers, and roots [Ruan et al., 1998]. Aharoni at al. [Aharoni et al., 2000] discovered a novel strawberry gene that plays a crucial role in flavor biogenesis in ripening fruit.

Complex gene expression patterns are being revealed with this new technology, involving large numbers of genes and unexpected components of cellular function in regulation and metabolism. As the complexity of microarray experiments increases, more sophisticated kinds of biological information can be extracted from microarray data. Enhanced exploitation of microarray technology requires more powerful and subtle data analysis and mining technology.

## Microarray Data Analysis

Several algorithms from data mining, machine learning, and parametric and non-parametric statistics have entered microarray data analysis [Aharoni et al., 2000,Seki et al., 2001,Wang et al., 1999,White et al., 2000]. Analysis techniques such as $k$-means clustering, clustering by principal components, average linkage clustering [Jain and Dubes, 1988], self-organizing maps [Golub et al., 1999], agglomerative and hierarchical algorithms [Eisen et al., 1998,Reymond et al., 2000], Bayesian methods [Smyth, 1996], plaid models [Lazzeroni and Owen, 2000], and support vector machines [Brown et al., 2000] have been featured in a majority of published research. A commonly accepted dichotomy of analysis techniques distinguishes between *supervised* and *unsupervised* methods.

## Unsupervised Methods

From a probabilistic viewpoint, unsupervised methods model the *unconditional* distribution (e.g., via densities) of gene expression data, as revealed by microarray experiments [Jordan and Bishop, 1997]. For example, the set of genes responding positively can be modeled by a functional form such as a Gaussian or as a mixture of functional forms (called a mixture model). The mixture model, in particular, uses superpositions of simpler densities (such as Gaussians defined with known means and covariance) to represent high-dimensional and high-variability regions of gene expression [Yeung et al., 2001]. The distributions modeled by these simpler densities can be viewed as *clusters*, which are categorically homogeneous subsets. Thus, clustering looks for regularities in the *training data* (the data resulting from microarray experiments) and can be used to provide a compact representation of the input problem domain. Different clustering methods have been proposed that represent clusters in different ways, using, for example, a representative exemplar of a cluster; a probability distribution over a space of attribute values; or necessary and sufficient conditions for cluster membership. To represent a cluster by a collection of training data and to `assign' new samples to existing clusters, some form of a utility measure is used. This is normally based on one or more mathematical properties -- such as distance, angle, curvature, symmetry, and intensity -- exhibited by the members of the cluster. $k$-means clustering, average linkage clustering, clustering by principal components, self-organizing maps, agglomerative and hierarchical algorithms are examples of this unsupervised mode of investigation.

## Supervised Methods

Supervised methods, on the other hand, can be viewed as modeling *conditional* distributions. Jordan and Bishop [Jordan and Bishop, 1997] demonstrate the value of viewing these techniques as modeling the dependence of a set of output variables on a set of input variables. For example, in addition to capturing the similarity between a set of positively-responding genes, supervised techniques can relate this gene expression to putative functional categorizations of the genes or other *a priori* knowledge. Thus, gene expression is an output variable, conditionally dependent on the input (such as functional category). Modeling the conditional distribution is important if we are interested in assigning (predicting) output values for new (untested) input value, as well as if we are interested in deriving phenomenological models that serve as initial explanations of the observations. Support vector machines [Brown et al., 2000], decision trees [Garofalakis et al., 2000], regression [Vapnik, 1998], discriminant analysis [Sullivan, 2001], and backpropagation neural networks [Gallant, 1993] are examples of supervised techniques.

## Model-based vs. Model-Free Methods

In addition, analysis techniques can be characterized as *model-free* (also called non-parametric, memory-based, or lazy) or as *model-based* (also called parametric or eager) techniques. A model-free technique makes no assumption concerning the underlying (conditional or unconditional) density function. Examples are nearest neighbor techniques and voting schemes. A model-based technique attempts to generalize from data by capturing a more succinct representation of data -- certainly more succinct than exhaustively enumerating the data. Formal methodologies such as the minimum description length (MDL) principle emphasize this view of learning as compression [Rissanen, 1978].

The importance of analysis techniques in finding coordinated sets of gene expression patterns is widely accepted [Schaffer et al., 2000]. As a result, researchers have spent considerable effort improving and tuning these techniques. Many clustering techniques are sensitive to the choice of the similarity metric used to distinguish intra-cluster variation from inter-cluster variation. Approaches such as permutation testing have been proposed to address this sensitivity for agglomerative and hierarchical clustering (K. Schlauch, personal communication).

While many of these techniques permit extremely efficient implementations, they suffer from two fundamental drawbacks:

1. An inability to incorporate prior biological knowledge. For example, clustering results are far from self-explanatory and a manual inspection is often required to decipher the biological implications of a set of genes being assigned to the same cluster, if this can be done at all. Supervised techniques studied in the microarray literature alleviate this problem to some extent but are limited by the second drawback.
2. The lack of expressiveness of the mined patterns; that is, each of these techniques can represent only a limited set of facts using its patterns. Furthermore, they can exploit prior knowledge in only a limited form, typically *a priori* functional categorizations.

The dichotomy between unsupervised and supervised methods has prompted several interesting observations. One particular strategy, studied by Sherlock [Sherlock, 2000] and Golub et al. [Golub et al., 1999] addresses the question "Is the overlap between the genes in a functional class and the genes in a particular expression cluster greater than would be expected by chance?" It is instructive to note that this has been viewed as a problem of correlating expression data to other information; thus many researchers [Cho et al., 1998, Kannan et al., 2001] have approached this problem by conducting supervised and unsupervised analyses as separate stages and later using statistical tests to determine the significance of such correlations.

## Environmental Effects on Tree Growth and Wood Properties

Forest trees are in the earliest stages of domestication. Both tree breeding and fundamental genetic studies have been greatly hindered by the long generation times and large size of forest trees. The technology of genomics and microarrays allows insights into the molecular basis of growth and specific wood properties without the need for extensive breeding experiments over many generations, and can be carried out on existing material or on young seedlings. Genomics allows insights into the physiology of otherwise intractable systems and allows us to benefit greatly from comparisons to model systems based on sequence information. The identification, functional analysis, and location of expressed genes in forest trees has many applications. One area of specific interest is the identification of the effects of abiotic stress on wood formation, particularly water stress, because of the effect of water stress on tree growth and on wood properties [Lev-Yadun and Sederoff, 2000,Costa et al., 1998,Costa and Plomion, 1999]. Growth and wood properties are important commercial factors because they affect the yield and quality of commercial forests.

A major motivation for the development of microarrays for forest trees is the potential for the analysis of environmental effects on trees in natural populations. The microarray system is potentially able to monitor acute and chronic environmental effects based upon the specificity and level of expression observed for a large suite of genes. The microarray method has great inherent accuracy and precision, but there is a great deal of development needed for this potential to be realized. One of the major barriers at present is the lack of appropriate computational tools for analyzing and interpreting the specific and the correlated effects on gene expression during development of adaptation to abiotic or biotic stress. Microarray technology can contribute greatly to studies of adaptation and response to climate change.

## Short- and Long-Term Adaptational Responses of Plants to Environmental Stress

The ability of a plant to protect itself against environmental stress is essential to its survival [Alscher et al., 1997]. Acclimation of plants to extreme environmental conditions or to rapid changes in growth conditions requires a global cellular response and changes in the expression of many genes. Exposure to extremes of light intensity and temperature, drought, and some herbicides can cause the downstream formation of reactive oxygen species (ROS). ROS may be present in the form of superoxide ($O_2^-$), hydrogen peroxide ($H_2O_2$), or the hydroxyl ion ($OH^-$). ROS, especially $OH^-$, are toxic because they can oxidize any macromolecule in the cell [Scandalios (ed.), 1997]. This potential threat to cellular function can cause protein unfolding, the inactivation of enzymes, DNA damage, mutation, lipid peroxidation, and consequent disruption of cell membrane function.

In animals, injury due to unchecked ROS damage has been linked to cancer and aging [Gilchrest and Bohr, 1997]. ROS have also been implicated as links in stress-responsive signal transduction mechanisms [May et al., 1998]. There is a suggested role for hydrogen peroxide in the signaling events leading to the activation of the defense-response [Mullineaux et al., 2000]. In some instances, the plant successfully adapts to its changed environment. Both short- and long-term effects occur. Shinozaki *et al.* [Shinozaki and Yamaguchi-Shinozaki, 2000] distinguish between rapid, emergency responses, and slower, adaptive responses associated with successfully attaining a new steady state under stress conditions.

Our understanding of genome-wide mechanisms contributing to successful adaptation in plant cells is incomplete. A coordinated global shift in gene expression in plant cells is expected to be involved in adjustment to unfavorable conditions. There is growing evidence that while the mechanism whereby a plant perceives and transduces a particular adverse environmental change is specific to that change, that there is considerable overlap among the mechanisms that respond to the class of adverse environmental changes [Uno et al., 2000] (see also Fig. 2). Genes controlling osmotic adjustment, protein stabilization, ROS detoxification, ion transport, membrane fluidity, gene activation, and signal transduction have all been implicated in stress adjustment responses in higher plants in separate experimental systems [Cushman and Bohnert, 2000, Shinozaki and Yamaguchi-Shinozaki, 1997]. Specifically, *dehydrins* have been shown to be associated with dehydration tolerance [Zhu et al., 2000]; *proteases* with signal perception and transduction [Callis and Vierstra, 2000]; *calmodulin-mediated processes and membrane transport* with salt tolerance [Geisler et al., 2000]; *cell wall extensibility events (extensins, proline-rich proteins)* with adaptation to drought stress [Wu and Cosgrove, 2000]; and *lignin production (laccases, phenylpropanoid pathway enzymes)* with growth under multiple stress conditions, including salt stress [Degenhardt and Gimmler, 2000] (see Fig. 6 and 7).

## A Search for Molecular Adaptation Mechanisms in Plants

Little information has been gathered for any experimental system that documents global changes in gene expression associated with successful, long-term adaptation to stress. This is in sharp contrast to the available data that documents response to short-term exposures. Although there are several molecular defense systems that respond to environmental stress in plants (Fig. 2), their relative importance for long-term stress resistance is not known. It is likely that the coordination, identity, timing, and level of induction or suppression of stress-responsive genes is critical for effective and sustained removal of toxic ROS. Redox sensing appears to play a central role in environmental stress responses. The effective repair and renewal of individual, stress-susceptible, macromolecules and associated cellular and physiological processes is essential for cell, tissue, and organism survival (see Fig. 2). Gasch *et al.* [Gasch et al., 2000] used microarray technology to investigate adaptive responses of the entire yeast genome to a series of abiotic stresses. They interpreted increases in transcript abundance in genes associated with signal transduction, chaperones, ROS detoxification, and bioenergetics as events involving essential processes to ensure cellular survival in the face of long-term oxidative stress.

## Choice of Experimental System

Comparison of the transcriptional responses of higher plants to environmental stresses is a powerful tool for understanding the functions of individual genes in the responses as well as the adaptive response of the genome as a system. The use of comparative functional genomics to study responses to environmental stress in forest trees enables new and different investigations of mechanisms of adaptation. Information on gene function related to environmental adaptation is far from complete even in the model plant Arabidopsis. Many genes have important interactions that may not be apparent from their primary function. Even when a homolog has been identified by comparative sequence analysis and a specific function has been implicated, many genes may have additional functions and interactions in a woody plant that might not be inferred from studies of Arabidopsis.

## Expresso: A Microarray Experiment Management System

Our research group has recognized the need to address all phases of a microarray experiment as a coherent whole and to fashion a computational system that integrates the design, analysis, and data management tasks as well as the laboratory and computational components. The Expresso system [Alscher et al., 2001] is designed to support all microarray activities including experiment design, data acquisition, image processing, statistical analysis, and data mining. Currently, the latter three stages are completely automated and integrated within our implementation. The data and physical flows realized using Expresso are shown in Fig. 1.

The design of Expresso underscores the importance of modeling both physical and computational flows through a pipeline to aid in biological model refinement and hypothesis generation. It provides for a constantly changing scenario (in terms of data, schema, and the nature of experiments conducted). The design, analysis, and data mining activities in microarray analysis are strongly interactive and iterative. Expresso thus utilizes a lightweight data model to intelligently "close the loop" and address both experiment design and data analysis. Expresso also uses inductive logic programming (ILP), a relational data mining technique to model interactions among genes and to evaluate and refine hypothesized gene regulatory networks. We refer the reader to [Alscher et al., 2001] for a more detailed exposition of the computational, algorithmic, and system implementation issues underlying the design of Expresso.

## Modeling With Expresso

The first step in modeling with Expresso entails defining semi-structured data records corresponding to information about material selection, PCR, randomization, spotting, hybridization, gridding, data extraction, and data analysis. An instance of all of these records thus represents a pipeline of stages involved in a single microarray experiment (a partial example is given in Fig. 3).

These self-describing descriptions serve several useful purposes. First, since they can be stored in a database and queried, programmatic descriptions of new experiments can be automatically created by writing queries. For example, a high-level specification such as "Design a new experiment using the layouts from 1999, the dye concentrations used by Mark, and the conditions of mild drought stress," or "Use the same experimental setup as EXPT-99-Pine-Drought, but with a signal threshold of 0.60" can be defined by writing database queries in languages like SQL and XML. In addition, biologists are able to interact with Expresso using abstractions such as stress experiments and expression levels, instead of the current emphasis on tedious details such as wells in microtitre plates or measured fluorescence in a tiff image. This can be achieved by providing an interface that masks how individual records are composed to arrive at full-fledged descriptions of experiments. Second, such descriptions can (optionally) be then used to manage the physical and computational execution environment (e.g., pipetting robots, image readers, and data mining software). For instance, it is possible to transform an Expresso description into the low-level programming code for controlling and driving laboratory instruments that have programmable interfaces supporting laboratory automation and management. Together, these features help us store descriptions, `run' the descriptions to obtain data, record the data back in the database, and associate the data with the description that corresponds to its experimental setup. Finally, having descriptions of experiments allows us to provide sophisticated services such as *change management*. For example, consider that two students configure Expresso independently with different choices for various stages in the pipeline and arrive at contradictory results. They could then query the database for "What is different between the experiments that produced data in directory A from the ones in directory B?" -- providing responses such as "The only difference is that a calibration threshold of 0.84 was used in B instead of 0.96 for A," which are obtained by automatically analyzing the descriptions.

In contrast to the variety of standards (many, XML-based) available for describing microarray data, our data model is thus aimed at capturing representations of experiments, not just experimental data. We posit that the description of an experiment is a more persistent representation of the data (it produces) than the data itself. As technology matures and evolves, recording how specific data was obtained is important for the purposes of ensuring repeatability and reliability. For example, if gridding technology improves, then `running' the same (stored) description with the new setup can be used to obtain new results.

One of the hallmarks of Expresso and its semi-structured representation is that the data model is lightweight and can elegantly adapt to changes in schema over time. There is nothing in our language design that commits us, for instance, to describing DYEs by two attributes (ref. Fig. 3). As new forms of DYEs are introduced into Expresso, the data model can `expand' to accommodate new fields and attributes, that were not applicable in older records. The design of the semi-structured language is beyond the scope of this article; a preliminary description is available in [Alscher et al., 2001]. Here, we specifically concentrate on using Expresso to understand stress responses in Loblolly Pine.

In this paper, we explicitly concentrate on the use of Expresso to mine global patterns of gene expression in order to uncover regulatory mechanisms that are essential for long-term adaptation to stress in woody species. In particular, our main focus is on gene expression associated with adaptation to drought stress over one growing season in loblolly pine.

# Results

In June 1999, drought stress was developed by withholding water from rooted cuttings of two unrelated loblolly genotypes from the Atlantic Coastal Plain, while control plants were watered normally. Mild and severe drought stress constituted pre-dawn water potentials (pressure bomb technique) of –10 to –12 bars, and –16 to –18 bars, respectively. Rooted cuttings were subjected to mild or severe drought stress for four (mild) or three (severe) cycles Plants were watered to field capacity once these pressure potentials were attained. Needles were harvested after the drought cycles were completed (adaptation). Mild stress produced little effect on growth with new flushes as in control trees. Imposition of severe stress resulted in growth retardation with markedly fewer new flushes compared to controls. Using RNA harvested from individual trees and different treatments we have determined global patterns of gene expression on microarrays. With algorithms incorporated into Expresso, we have identified genes and groups of genes involved in stress responses.

## Effect of Mild and Severe Stress on Gene Expression

Data were obtained for the control versus mild stress condition and for the control versus severe, non-adaptive, condition for Genotype D and for the control versus mild condition alone for Genotype C. We performed preliminary statistical analysis on both genotypes but applied the inductive logic programming technique only to data from Genotype D. The reason for this restriction involves the theoretical model of machine learning assumed by typical ILP implementations. All expression data are distilled within Expresso into a single metric of up-expressed, down-expressed, or unchanged. Inclusion of Genotype C into our study would imply that the probability distribution from which examples (instances of gene expression data) are picked is nonuniform (ref. the control versus mild condition). However, currently available ILP systems do not provide systematic ways to incorporate this prior knowledge (of the distribution of examples) as learning parameters. While theoretical analyses are definitely feasible, our goal was to ensure the biological validity of the hypotheses generated. A total of 37 rules were mined by ILP for Genotype D.

Using Expresso, we determined that 72 of the 384 cDNAs present on the microarray showed an increase in transcript abundance relative to controls at the end of four repeated cycles of exposure to mild stress in Genotype D. Of those 72 cDNAs, 69 showed either a decrease or no change in the control versus severe comparison. Expresso mined this observation as the rule:

$$\texttt{\~{}level(A,CvsS,positive) :- level(A,CvsM,positive).} \qquad (1)$$

In rules, the notation `~` means logical negation or `not'. Hence, this rule means that if a clone (`A`) is up-expressed in the `CvsM` (control versus mild drought stress) comparison, then it was *not* up-expressed in the `CvsS` (control versus severe drought stress) comparison. This rule is supported by observations of transcript abundance of 69 out of 72 relevant cDNAs. This gives a confidence level of 69/72 (about 96%).

These clones, and their associated functional categories, are candidates for drought stress adaptation genes and for participation in associated mechanisms. The functional categories of the positive responders in the `protective processes' grouping are shown in Fig. 7. The remaining 257 cDNAs were either unaffected (204), or showed a decrease relative to the controls (43). Among the positive responders were genes already associated with water stress responses such as the dehydrins and water channel proteins or aquaporins. The class of transport proteins, into which the aquaporins fall, showed a negative response in the mild versus severe stress comparison, and a positive in the control versus mild contrast. This was indicated by the next two rules (confidences 63.63% and 66.67% respectively):

```
level(A,CvsM,positive) :- category(A,membranetransportprotein).     (2)

level(A,MvsS,negative) :- category(A,membranetransportprotein).     (3)
```

In the case of both dehydrins and aquaporins, different cDNAs responded to probes from the two genotypes. Aquaporins associated with both the tonoplast and the plasma membrane were present on the array. cDNAs representing both subcellular locations responded positively in the control versus mild stress comparison, suggesting the importance of water channel proteins in the tonoplast and the plasma membrane for adaptation to mild drought stress. Genes encoding heat shock proteins (HSPs) -- HSP70 (chloroplast-associated chaperone function [Rial et al., 2000]), HSP23 (LEA-like genes [Dong and Dunstan, 1996]), and HSP100 (thermotolerance [Hong and Vierling, 2000]) -- also responded positively to mild drought stress (confidence 83.33%):

```
level(A,CvsM,positive) :- category(A,heat).     (4)
```

In contrast, HSP80s (thought to be involved in chromatin organization [Schnaider et al., 1999]) did not respond in either genotype. In some cases, different cDNAs responded to probes from the two genotypes (see Fig. 4 and Table 1 for a summary of results obtained for HSPs). Several HSPs (Clone 226, an HSP 101; clone 228, an HSP 23.5; and clone 296, an HSP 70) showed a positive response in the control versus mild comparison and a negative in the control versus severe comparison, making them strong candidates for drought stress adaptation genes. Rubisco-binding proteins were also among the positive responders in the control versus mild drought stress comparison. In contrast, Rubisco-binding proteins were unchanged in the mild versus severe comparison, suggesting that these genes may not be among the class of candidates for stress adaptation. LP-3, an established water-stress inducible gene in loblolly pine, responded positively to probes from both genotypes. No difference in expression level of LP-3 was detected in the control versus severe contrast in Genotype D. LP-3, a protein with a chaperone function, therefore falls into the class of candidate genes associated with resistance to mild drought stress. There was no detectable difference among the thiol-utilizing enzymes for the control versus mild, or the control versus severe stress comparisons. There is much documentation demonstrating the involvement of thiol-utilizing enzymes in short term responses to oxidative stress, but little to document events associated with long-term adaptation.

The class of genes categorized very loosely as ``isoflavone reductases,'' of which four separate ESTs were included on the array, exhibited positive responses in both genotypes in the control versus mild drought stress comparison, with two ESTs with greatest resemblance to phenylcoumarinbenzylic ether reductases responding in Genotype C, and two ESTs corresponding to the closely related pinoresinol-lariciresinol reductases in Genotype D. On the other hand, the IFR homologs showed no detectable change in the mild versus severe comparison, suggesting a response to stress, but no correlation with successful adaptation to mild drought conditions. Genes associated with lignin biosynthesis also responded positively, as did GST, proteases, and receptor-like protein kinases. In the case of cell wall associated genes, positive change was detected in the control versus mild comparison (confidence: 81.25%)

$$\text{level(A,CvsM,positive) :- category(A,cellwallrelated).} \qquad (5)$$

and a negative response for lignin biosynthesis genes in the control versus severe contrast (confidence: 81.81%)

$$\text{level(A,CvsS,negative) :- category(A,ligninbiosynthesis).} \qquad (6)$$

suggesting a role for lignin biosynthesis in drought stress adaptation.

# Discussion

Using cDNA microarrays, we have investigated expression patterns of genes in needles of loblolly rooted cuttings (equivalent in size and development to one-year-old seedlings, but of identical genotype) from two different unrelated genotypes from the Atlantic Coast Plain that had successfully adapted to cycles of mild drought conditions over a growing season. We have compared those results with results obtained from rooted cuttings of one of the genotypes exposed to more severe, non-adaptive, conditions over the same time period.

The expression data reported here reflect the adaptational adjustments made by loblolly pine needles to long-term and intermittent drought stress. The control versus mild stress comparison for two unrelated genotypes identifies candidate functional categories for drought stress tolerance and resistance. The positive response of LP3 in the control versus mild stress comparison, a known water-stress inducible gene in *Pinus taeda,* serves as a positive control for our data. Dehydrins and aquaporins are among the responders, as would be expected from their established physiological roles. There were many aquaporin ESTs in our microarray group. However, we cannot definitively distinguish between ESTs from one gene or from closely related members of a multi-gene family. The aquaporins were divided among tonoplast and cell membrane-associated groups; thus we are dealing with at least two different genes. The genes from the heat shock proteins that responded fell into three groups. Of these three, two -- the HSP70s and the HSP23s -- have known chaperone functions, not necessarily related to heat shock responses, and can perhaps be regarded as fulfilling a maintenance or repair role in cells that are coping with mild drought stress on an ongoing basis. The positive response of the HSP70s, which fulfill a chaperone or targeting function for proteins synthesized in the cytosol and destined for the chloroplast, is in agreement with the increases in transcript abundance of photosynthesis-associated genes. These HSPs showed no difference, or were negative in the control versus severe stress comparison. Thus, the HSPs are candidate genes for drought resistance.

The HSP100s, which also showed a positive response in the control versus mild stress comparison, are more definitively associated with heat shock itself. Their response may indicate the existence of a common core of genes that respond to a range of different stresses; a result that is in agreement with many others in the literature. The IFR homologs that responded positively in both genotypes, as well as the GSTs, may also fall into the class of a core of stress responsive genes, although not in the class of stress resistance genes *per se*. Both the IFR homologs and the GSTs are most commonly associated with responses to biotic and xenobiotic stress and not to the abiotic challenge presented by drought stress. Response to increased ROS levels may be the common denominator for these changes in the various functional categories.

Our results present a snapshot of the state of gene expression in loblolly needle tissue that has adapted to mild drought stress. A detailed time course study is needed to identify events in gene expression that lead to adaptation. Many, transitory, stages in signal perception and transduction can only be captured by sampling early on in the adaptation process. We plan to set up a sampling scheme to glean evidence for the physiological changes that underlie short-term emergency adjustments and to identify those changes that are essential for subsequent, long-term adaptation.

These requirements point to the future directions in the development of Expresso. We are now extending Expresso to intelligently integrate experiment design and data analysis. This will provide us the ability to use run-time information from the results of data mining to make recommendations about the earlier stages in experimentation, such as layout, randomization, and choice of clones for the next iteration of studies.

A long-term biological goal is the modeling of the dynamics of adaptation to environmental changes. Understanding the qualitative and quantitative responses of metabolic pathways to external and internal signals implies the need to integrate biological knowledge drawn from gene expression studies together with information from proteomics and metabolic profiling. The data management architecture of Expresso is designed to have the flexibility to support this aspect of inquiry.

## Detection of Gene Expression Changes

The inevitable presence of experimental errors complicates the determination, for each clone (cDNA) represented on a set of microarrays, whether that clone shows clear changes in transcript levels under the experimental conditions considered. Various statistical tests suggested in the literature do not fully utilize the available information or make assumptions that are probably too strong and unrealistic. Chen et al. [Chen et al., 1997] derive a probability density of transcription ratios under strong (and highly unrealistic) distributional assumptions. They investigate neither the effects of deviations from their strong assumptions nor all possible sources of variation (e.g., due to background estimation). Some authors (e.g., Claverie [Claverie, 1999]) have suggested the use of t-tests applied to intensity values. This approach requires replication of the same gene on one or more arrays and the use of paired t-tests or non-parametric paired tests such as the sign test. These tests are expected to have poor efficiency with few replications. Hilsenbeck et al. [Hilsenbeck et al., 1999] uses prediction regions from principal components, while Greller and Tobin [Greller and Tobin, 1999] describe a decision function using a statistical discordancy test. Several approaches involving Bayesian methods have also been proposed. One is the Hierarchical Generalized Linear Model (see Daniels et al. [Daniels and Gatsonis, 1999] for general methodology and Lee et al. [Lee et al., 2000] for application to activity data), which can model and estimate noise variance components if replication is available.

The methodology followed in Expresso is qualitatively different; we obtain multiple (typically 16) log-calibrated-ratios for a single replicated clone; by observation, we find that the log-calibrated-ratios for a single clone do **not** follow a normal distribution. Far from it, each distribution is spread relatively evenly over a large range. Statistical analysis based on mean and standard deviation will thus be overly pessimistic in identifying clones that are up- or down-expressed. Given this observation, we make a much weaker probabilistic assumption on the distribution; we assume that a clone whose expression is not different between a probe pair will show a distribution centered at a mean log ratio of 0.0. Our assumption of a zero-centered distribution is more general than the assumption of a particular distribution, such as a normal distribution, and hence is more likely to hold in a real experiment. In a zero-centered distribution, the probability that any particular log ratio is positive (or negative) is 0.5. The number of positive (or negative) log ratios follows a binomial distribution with parameters 16 and 0.5. The probability of 12 positive log ratios (or 12 negative log ratios), out of 16, for a clone whose expression was unaffected by drought stress is 0.0384064. Consequently, a clone with 12 or more positive log ratios is up-expressed with a probability of 0.96. Our more general assumption avoids the trap of having to classify the response of each **spot**; rather we classify the response of each **EST** as one of: up-regulated, down-regulated; or no clear change. Our three-way response classification allows us to develop meaningful relationships among genes and among treatments and also provides sufficient results for the use of sophisticated data mining techniques (see below).

## Inductive Logic Programming

Our analysis methodology is motivated by the need to connect functional categorizations of genes with systematic variations in expression levels. In his review of expression data analysis, Sherlock [Sherlock, 2000] discusses two techniques for correlating biological information with expression data. The first technique builds a two-way classification predictor based on weighted votes provided by gene expression from tissues in each of the two classifications. Golub, et al. [Golub et al., 1999] successfully apply the technique to classifying leukemia type from human tissues. The second technique first organizes genes into clusters using k-means clustering of expression data and then computes statistical correlations between each cluster and each of a set of functional categories. Our analysis methodology is fundamentally different from the techniques discussed by Sherlock [Sherlock, 2000]. We use the inductive logic programming (ILP) approach [Muggleton and Feng, 1990, Muggleton, 1999, Dzeroski, 1996] as an aid in data mining and formation of biological hypotheses. ILP is a technique that provides, in one integrated procedure,

1. A way to correlate output variables (gene expression) with input variables (functional categorizations, for instance);
2. A richer representational basis (allowing the incorporation of expressive *a priori* biological knowledge, not limited to functional categories); and
3. A methodology of abduction, a process of hypothesis formation (that can involve finding coordinated sets of gene expression data).

ILP takes input data expressed as gene expression levels in particular experiments, relationships between experiments, functional categories, and any biological knowledge that is available. As output, it provides rules of the form

```
level(A,CvsS,negative) :- level(A,CvsM,positive).
```

where C, M, and S represent control, mild drought stress, and severe drought stress conditions, respectively. This rule states: ``If a clone (represented in the rule as `A`) was positively expressed in the control versus mild drought stress comparison (`CvsM`), then it was negatively expressed in the control versus severe drought stress (`CvsS`) comparison.'' The restated rule is easily understood and can be used in later diagnostics and what-if analyses. ILP algorithms do not require explicit invocation or instructions to mine rules across comparisons. Rules are produced as the result of a process of systematic search for succinct, conceptual clusters of data. ILP can thus be used to find patterns within a given comparison, across comparisons, and across functional categories. In contrast, a purely unsupervised method may recognize a particular group of clones in the control versus mild drought stress comparison that exhibit a positive response and also recognize another group of clones in the control versus severe drought stress comparison that exhibit a negative response. However, *it cannot model the connection between these two comparisons* (unless we know beforehand that this kind of connection is what we are looking for). A purely supervised technique can make such a connection *only after the above clusters are modeled (recognized and given as input).* **ILP subsumes both these modes of analysis in rule formation.** In addition, ILP rules can be recursive, a feature that makes them amenable to discovery of complex relationships involving hierarchies of functional categories [Muggleton, 1999] (see results below).

## Data preparation for ILP

ILP systems typically take a database of positive examples (correct gene expression data), negative examples (information known *not* to represent correct gene expression data) and background knowledge (functional categories, for example) and attempt to construct a predicate logic formula (such as `level(A,B,C)`) so that all (most) positive examples can be logically derived from the background knowledge and no (few) negative examples can be logically derived. The need for negative examples can be seen by observing that ILP algorithms conduct mining by searching through a space of possible patterns. Such a space is typically organized as a specialization-generalization hierarchy. The more specific patterns are at the bottom of the subsumption lattice and the most general patterns are at the top of the lattice. While ILP algorithms differ in how they navigate, prune, or focus on this space of patterns, all of them require a way to *evaluate* any specific pattern encountered in such a search. One useful form of such evaluation pertains to how accurately the pattern fits the positive examples and how accurately it fails to fit the negative examples. The more negative examples that are *covered* by a pattern, the less likely that it will be a good representation (or predictor) for the underlying data distribution. Hence negative examples are important to ensure that the mining does not produce *overly general* patterns. For instance, suppose that the only clones presented to ILP are from the `heat' category and that they all responded positively in a certain comparison. Mining that *all clones* respond positively in that comparison would certainly be a valid (from the data distribution viewpoint) but dangerous (from the biological viewpoint) conclusion to make. If additional clones (from other categories) are presented that do not respond positively in the given comparison, then data mining can correctly infer that it is the membership in the `heat' category that co-occurs with the positive expression. Similarly, negative examples help produce valid patterns by defining the boundaries of generalization without any truly ``additional'' information.

A partial listing of our database tuples is shown in Fig. 5. Negative examples are automatically generated by invoking the closed-world assumption, which states that all relevant facts are stored in the tables and facts not recorded can be taken to be false. For tables that have a large number of columns (high arity), this might cause us to generate a huge number of negative examples. The typical solutions are to (i) place restrictions on how variables are coupled in an ILP rule (allowing us to use them to generate negative examples), (ii) perform a probabilistic analysis by constructing so-called *stochastic logic programs*. In our experiments, negative examples are easy to generate since the only variability allowed in the `level` table is in the `Expression` column. Thus, if a clone was positive for a particular comparison, we can declare two negative examples, namely that it was negative and unchanged for the (same) comparison. For our experiments, we made use of the Aleph ILP software [Srinivasan, 2001] from the Oxford University Machine Learning Laboratory.

## Attribute-value versus Relational Learning Techniques

ILP is a *relational learning* technique, distinguishing its representational basis from those of *attribute-value* based techniques (discussed in the beginning of this section). Formally, ILP uses a representation of (a proper subset of) first-order predicate logic, whereas attribute-value techniques work at the level of propositional logic. Its expressiveness makes it a highly desirable tool in structured domains (such as microarray data analysis) where comprehension and interpretation of patterns is important. Notice that to achieve similar results with attribute-value techniques requires (i) explicit enumeration of all possible types of rules; (ii) data preparation for each type of rule; and (iii) recasting the rules in biological terms. For even three comparisons and 100 functional categories of ESTs, there are potentially $2^{100}$ x $2^3$ possible types of rules. ILP techniques use pruning algorithms that focus on the most promising scenarios. Obtaining the same effect manually with attribute-value techniques is practically impossible because of the huge number of types of rules.

ILP techniques can incorporate prior biological knowledge; for example, if the biologist knows that there is a possible connection between protective processes such as ROS detoxification and protected processes such as the reductive pentose phosphate pathway, an ILP execution can be modeled to exploit such knowledge. In many cases, such prior knowledge is also helpful in speeding up ILP.

# Methods

## Plant Material

Rooted cuttings of loblolly pine equivalent in size to one-year-old seedlings were obtained from Dr. Barry Goldfarb at NCSU and were cloned from two unrelated genotypes (clones C and D) from the Atlantic Coastal Plain provenance.

## Choice of Target cDNAs

A substantial number of expressed genes (approximately 15,000) have now been identified from loblolly pine as part of the Pine Genome Sequencing Project [NCSU Forest Biotechnology Group, 2001]. The predominant source of these expressed genes is from wood forming tissues. These tissues are rich in expressed genes involved in cell wall biosynthesis and in intracellular signalling. Microarrays have been used with a small subset of these expressed genes to examine gene expression during development and under environmental stress. The long term goals of this project are to identify the genes expressed during wood formation, to identify the time and place of their expression at the cellular level, and to correlate the effects of their expression with variation in wood properties. This approach depends on genetic mapping of expressed genes and the correlation of the map positions of loci affecting quantitative variation in wood properties with the location of specific expressed genes.

Of the ESTs sequenced by the Pine Genome Sequencing Project, many have a proposed functional annotation derived from a BLAST search of protein databases. From this annotation, we selected 384 ESTs from differentiating xylem or shoot tips representing genes of annotated function. We grouped the 384 pine ESTs into four major functional categories (as shown in Fig. 6) -- gene expression, signal transduction, protective processes, and protected processes. A complete list of our functional categories and groupings is available at

    http://bioinformatics.cs.vt.edu/~ralscher/functional_categories.html,

and a detailed list, including annotations for individual clones, at

        http://bioinformatics.cs.vt.edu/~ralscher/clones_annotation.html.

Our selection of clones includes genes involved in proposed resistance processes and signal transduction mechanisms, as well as genes associated with processes expected to be vulnerable to drought stress such as those associated with carbon metabolism, photosynthesis, and respiration. Genes associated with ROS detoxification, cell wall extension, lignin biosynthesis, and chaperone function as well as stress specific genes, such as aquaporins and dehydrins were included in the protective processes category (see Fig. 7). We also included genes known to respond to other stresses, such as UV irradiation, pathogen invasion, and sulfur stress (`isoflavone reductase-like') [Gang et al., 1999] and xenobiotic stress (glutathione-S-transferases) (see Figs. 6 and 7).

## PCR Amplification

384 ESTs from differentiating xylem or shoot tips [NCSU Forest Biotechnology Group, 2001] with putative functions of interest were selected and PCR amplified using M13 forward and reverse universal primers. PCR was performed in a 50 μl reaction containing 39.1 μl ddH$_2$O, 5 μl 10x PCR buffer, 1.5 μl MgCl$_2$ (50mM), 1 μl dNTPs (lOmM each), 1 μl M13 forward primer (10 μM), 1 μl M13 reverse primer (10 μM), 0.4 μl TAQ polymerase (5U/μl), and 1 μl cDNA diluent. Amplifications were carried out in a MJ Research thermocycler (Waltham, MA, USA). Denaturation was performed at 94° C for 30 sec, followed by primer annealing at 57° C for 1 min. Chain elongation took place at 72° C for 4 min. These steps were repeated for 35 cycles. Final chain elongation took place at 72° C for 10 min. PCR products were then electrophoresed in 1.5% agarose gels, stained with ethidium bromide, and visualized using UV light. This step was necessary to confirm both quantity and quality of the PCR reactants.

## Microarray Design and Layout

The 384 ESTs were organized in 4 microtitre *source plates* after PCR amplification. Expresso generated a design for printing two types of microarrays that replicated each clone four times in each microarray type and that placed the replicates at random locations in a microarray so that any systematic errors due to location can be analyzed and corrected (see Fig. 8). Implementing a replicated, randomized design required the use of two robots. First, the contents of the 4 source plates were re-pipetted into 8 sets of 4 microtitre *printing plates,* using a TECAN Genesis 2 robot; each set was an *independent randomization* of the 4 source plates. Second, slides were printed using a Stanford-type arrayer (see `http://cmgm.stanford.edu/pbrown/mguide`) built in-house at NCSU. The randomized design guaranteed that every EST was represented once in each set of 4 microtitre plates and that each set was a different physical arrangement (permutation) of the 384 ESTs. Two types of microarrays, A and B, were printed; each had 4 replicates of the 384 ESTs, but the random arrangement of clones differed in the two microarrays. Consequently, each array type has 4 replicates of each EST, randomly placed, and a total of 1536 spots. Each glass slide contained 2 identical arrays (either type A or type B); therefore, each slide had a total of 8 replicates of each EST.

## Slide Preparation

After printing, slides were processed according to the manufacturer's instructions (Telechem International, Sunnyvale, CA), with some modifications. In the first step, slides were first rinsed in 0.2% SDS twice for two minutes, with vigorous agitation; were then rinsed in distilled water twice for two minutes; and were finally transferred to boiling water for two minutes and cooled to room temperature for five minutes. In the second step, 1.5 g sodium borohydride was dissolved in 450 ml phosphate buffered saline to which 133 ml of 100% ethanol was added immediately prior to use. In the third step, the slides were first transferred to the sodium borohydride solution for five minutes; were then rinsed in 0.2% SDS for one minute three times; and were finally rinsed once in distilled water for one minute. Array boundaries were marked with a diamond-tipped pen. In the final step, the slides were dried by centrifugation and stored in the dark in a dessicator at room temperature.

## Hybridization

Each comparison of treatments (control versus mild drought stress; control versus severe drought stress; and mild versus severe drought stress) was done with hybridizations involving 4 microarrays on 2 slides of different types, comprising a total of 16 replicates of each EST. Total RNA was isolated from 11 separate needle samples harvested at the end of the growing season by the method of Chang et al. [Chang et al., 1993], and was the source of cDNA to probe the microarrays, after processing using the Genisphere Gene MicroExpression Kit. The two arrays present on each slide were hybridized with probe pairs labeled reciprocally with Cy3 and Cy5 dyes. A ScanArray 4000 was used to scan the slides after hybridization (Packard BioScience). The resultant TIFF images were analyzed initially using MicroArray Suite (Scanalytics, Fairfax, VA).

## Image Processing

For each cDNA spot, a *calibrated ratio* of intensities in the Cy3 and Cy5 channels was calculated and subsequently analyzed. We used Microarray Suite by Scanalytics to calculate these calibrated ratios. The techniques used by Microarray Suite are described in Chen *et al.* [Chen et al., 1997]. In general, we applied the default values used by Microarray Suite during our processing. A grid was placed manually on the TIFF images, identifying the locations of cDNA spots. For each grid location, the pixels comprising the spot contained within were determined using the Mann-Whitney statistical test at 99% confidence level. All remaining pixels comprise the background surrounding the spot. The mean intensities of the spot pixels and background pixels are calculated for each channel, from which a background-corrected ratio is computed. Finally, a calibrated ratio for each spot is computed used the iterative method described by Chen *et al.* [Chen et al., 1997].

# Acknowledgments

This work was supported in part by National Science Foundation Award number 9975806 to Ronald R. Sederoff entitled *Genomics of Wood Formation in Loblolly Pine.*

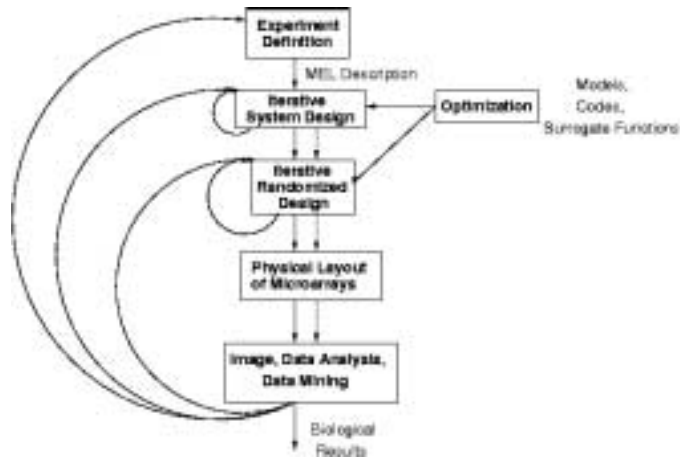

**Figure 1:** Execution and flows in Expresso. The solid lines indicate computational flows; the dashed lines indicate realizations of the pipeline in devices such as robots and image readers.

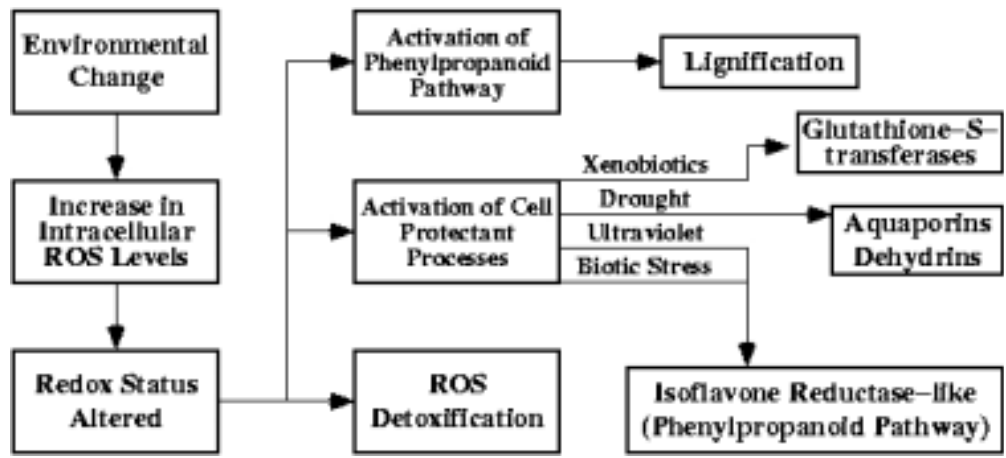

**Figure 2:** A Scenario for Specific and General Stress Responses in Plant Cells. Upon imposition of oxidative stress, ROS levels increase, and cellular redox-sensing mechanisms are activated. General downstream events include the activation of ROS detoxification mechanisms, such as the ascorbate glutathione scavenging cycle. Events specific to individual stresses include the activation of aquaporins in response to drought stress, and the activation of isoflavone reductase-like genes in response to pathogen invasion or UV-irradiation.

```
EXPERIMENT PINE_DROUGHT_GROWTH May-August,2000 "384 clones"
...
DYE CY3 "Genisphere Kit"
DYE CY5 "Genisphere Kit"
...
PRINTING_ROBOT NCSU_FBC "Brown-type robot at NCSU"
...
PRINTING_CONFIGURATION Stanford4x16x24   4 16 24 QUADRANTS
PRINTING_CONFIGURATION Stanford4x22x24   4 22 24 QUADRANTS
...
TISSUE D4M      D4      Needles Unstressed (Control)
TISSUE D4I      D4      Needles Intermediate Stressed
...
```

**Figure 3:** Example description of a microarray experiment in Expresso which includes the name of the experiment, the dyes used, description of the printing robot, printing formats, and the tissue configurations. The lines are in a self-describing format where the start field identifies the nature of information modeled.

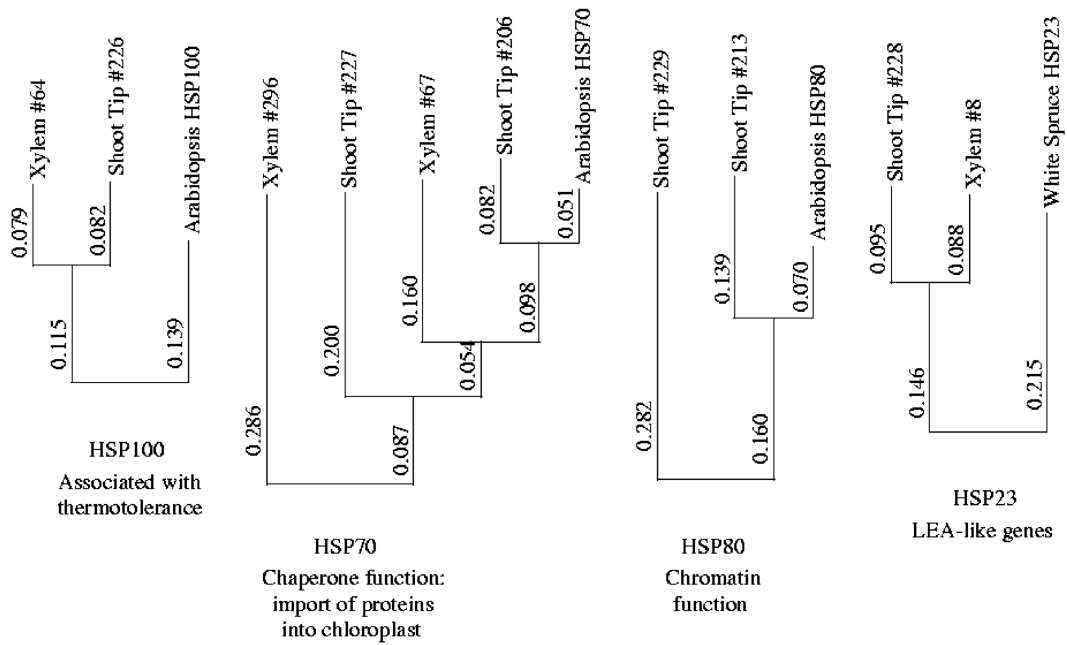

**Figure 4:** A dendrogram obtained using Expresso, ClustalW, and njplot showing the relationship of known HSP proteins of cDNAs among the EST stress set. ESTs are identified on the dendrogram by their origin and the number assigned to them in the microarray project. The corresponding Arabidopsis HSP sequence is included in each case.

| level | | | | category | |
|---|---|---|---|---|---|
| CloneID | Comparison | Expression | | CloneID | Category Name |
| … | … | … | | … | … |
| 4 | `CvsM` | positive | | 5 | RPPP |
| 5 | `CvsM` | positive | | 5 | Carbon Metabolism |
| 20 | `CvsM` | negative | | 7 | Thiol-Utilizing Enzymes |
| … | … | … | | 8 | Heat |
| 7 | `CvsS` | positive | | 20 | Drought Stress Responsive |
| 8 | `CvsS` | negative | | … | … |

```
category(X,Environment) :- category(X,Heat).
     category(X,Carbon Metabolism) :-
              category(X,RPPP).
```

**Figure 5:** Input database design for inductive logic programming (ILP). The **level** table contains information about the expression levels of individual clones, for all comparisons. It constitutes the positive examples. The **category** table records available functional classifications of all clones. Background knowledge consists of category containment relations, e.g., ``any clone that is classified under the `heat' category also belongs to the `environment' category.'' The negative examples are not shown. RPPP is the reductive pentose phosphate pathway.

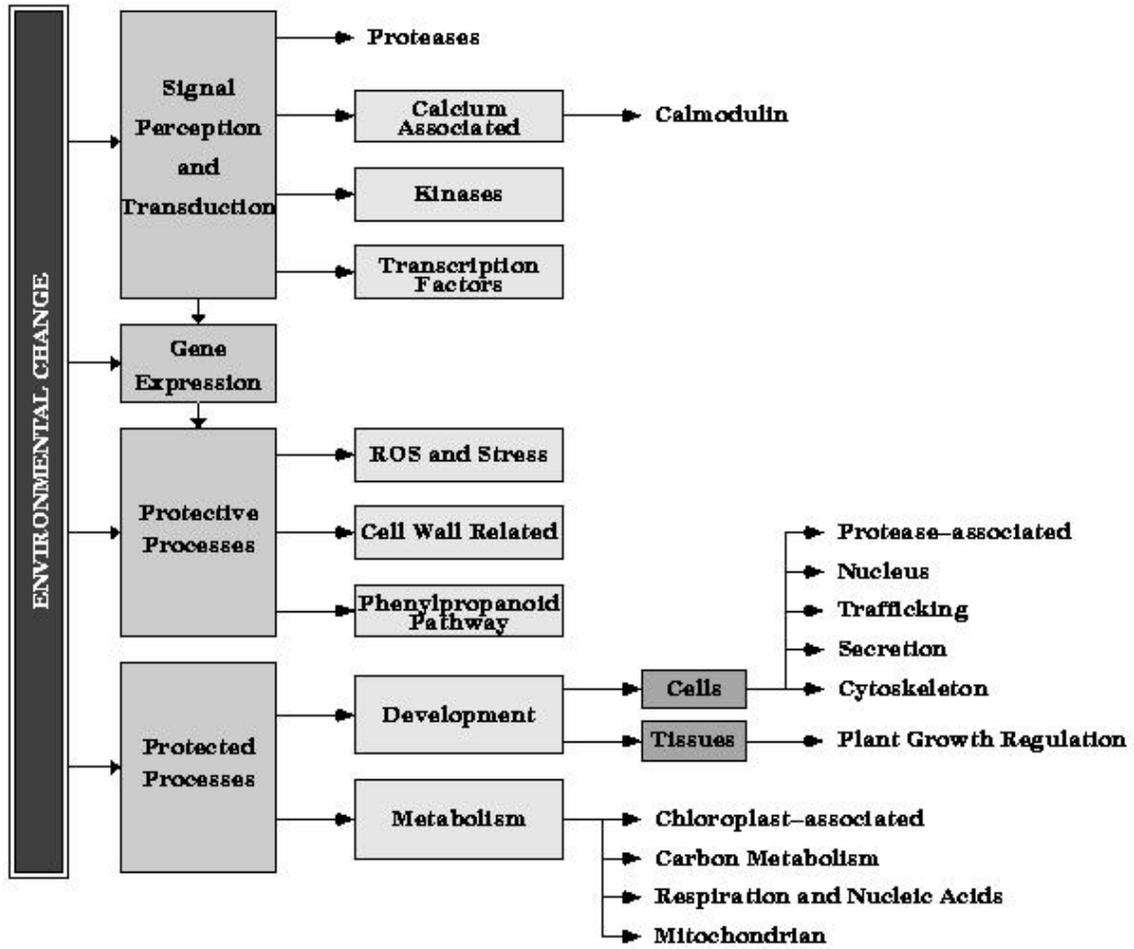

**Figure 6:** Categories and Groupings in Development and Metabolism.

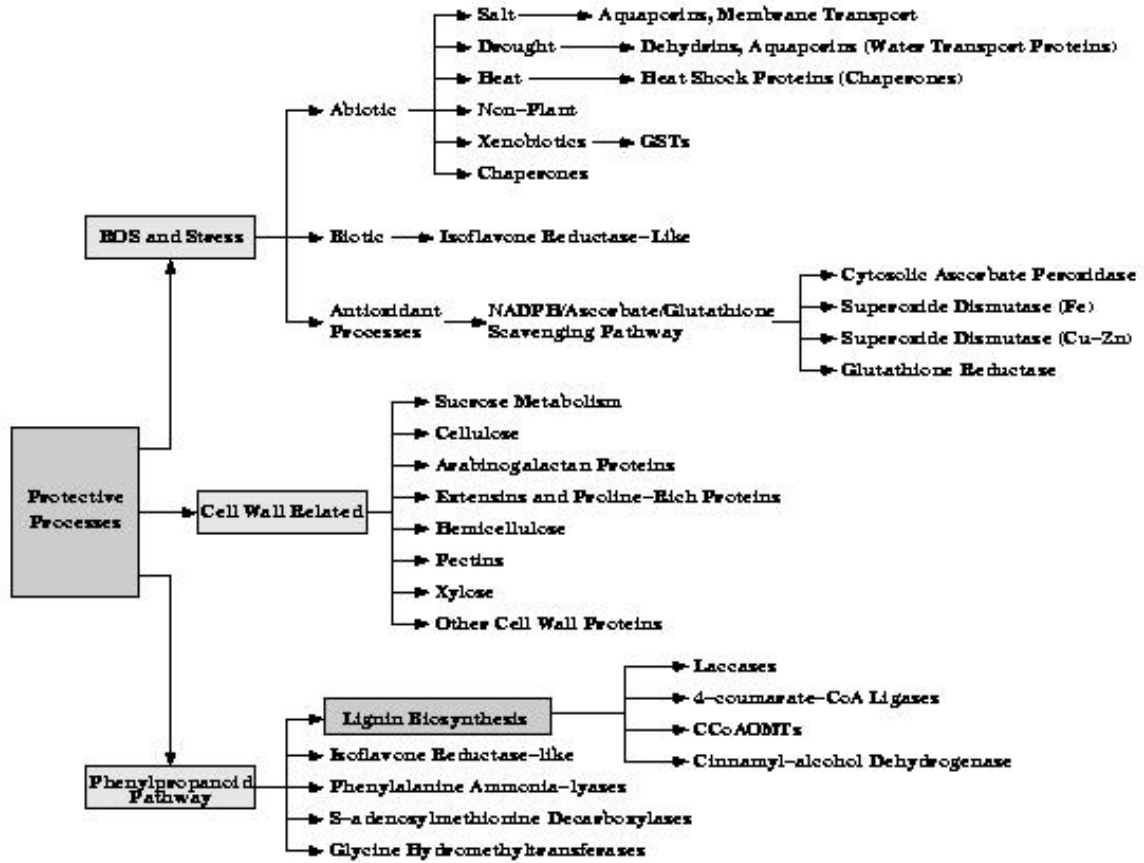

**Figure 7:** Categories and Groupings within Protective Processes.

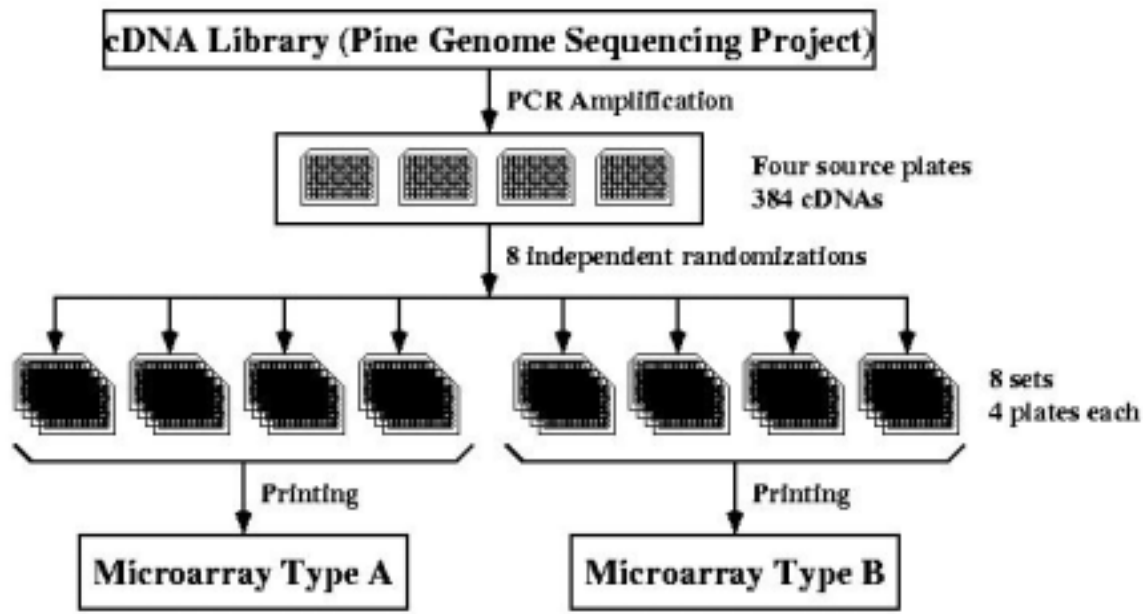

**Figure 8:** Design of Microarrays in Expresso. Slides containing the selected 384 cDNAs were printed as indicated above.

**Table 1:** Heat Shock Protein Responses Among the EST Stress Set in Genotypes C and D. Control versus Mild Drought Stress Comparison.

| HSP Type | EST Origin | Genotype C | Genotype D |
|----------|------------|------------|------------|
| HSP 23 | Xylem #8 | + | 0 |
|  | Shoot Tip #228 | + | + |
| HSP80 | Shoot Tip #213 | 0 | 0 |
|  | Shoot Tip #229 | 0 | 0 |
| HSP70 | Shoot Tip #206 | + | - |
|  | Xylem #67 | 0 | + |
|  | Shoot Tip #227 | 0 | 0 |
|  | Xylem #296 | + | + |
| HSP100 | Shoot Tip #226 | 0 | + |
|  | Xylem #64 | + | + |